\documentclass[journal=jpccck,manuscript=article]{achemso}

\usepackage{chemformula} 
\usepackage[T1]{fontenc} 

\usepackage{amsmath,txfonts,bm}
\usepackage{siunitx}
\usepackage{color}
\usepackage{ulem}
\usepackage{comment}

\author{Kazutaka Nishiguchi}
\affiliation[Kobe University]{Graduate School of System Informatics, Kobe University, 1-1 Rokkodai-cho, Nada-ku, Kobe 657-8501, Japan}
\alsoaffiliation[Fujitsu]{Fujitsu Research, Fujitsu Limited, 4-1-1, Kamiodanaka, Nakahara-Ku, Kawasaki, Kanagawa 211-8588, Japan}
\email{k.nishiguchi@fujitsu.com}
\author{Ryota Yamamoto}
\affiliation[Kobe University]{Graduate School of System Informatics, Kobe University, 1-1 Rokkodai-cho, Nada-ku, Kobe 657-8501, Japan}
\author{Meguru Yamazaki}
\affiliation[Fujitsu]{Fujitsu Research, Fujitsu Limited, 4-1-1, Kamiodanaka, Nakahara-Ku, Kawasaki, Kanagawa 211-8588, Japan}
\author{Naoki Matsumura}
\affiliation[Fujitsu]{Fujitsu Research, Fujitsu Limited, 4-1-1, Kamiodanaka, Nakahara-Ku, Kawasaki, Kanagawa 211-8588, Japan}
\author{Yuta Yoshimoto}
\affiliation[Fujitsu]{Fujitsu Research, Fujitsu Limited, 4-1-1, Kamiodanaka, Nakahara-Ku, Kawasaki, Kanagawa 211-8588, Japan}
\author{Seiichiro L. Ten-no}
\affiliation[Kobe University]
{Graduate School of System Informatics, Kobe University, 1-1 Rokkodai-cho, Nada-ku, Kobe 657-8501, Japan}
\author{Yasufumi Sakai}
\affiliation[Fujitsu]{Fujitsu Research, Fujitsu Limited, 4-1-1, Kamiodanaka, Nakahara-Ku, Kawasaki, Kanagawa 211-8588, Japan}
\title{Molecular Dynamics Simulations of SrTiO$_3$ with Oxygen Vacancies using Neural Network Potentials}

\begin{document}

%
%
%
%
%

\begin{abstract}

A precise analysis of point defects in solids requires accurate molecular dynamics (MD) simulations of large-scale systems. 
However, ab initio MD simulations based on density functional theory (DFT) incur high computational cost, while classical MD simulations lack accuracy. 
We perform MD simulations using a neural network potential (NNP) model (NNP-MD) to predict the physical quantities of both pristine SrTiO$_3$ and SrTiO$_3$ in the presence of oxygen vacancies (V$_{\text{O}}$). 
To verify the accuracy of the NNP models trained on different data sets, 
their NNP-MD predictions are compared with the results obtained from DFT calculations. 
The predictions of the total energy show good agreement with the DFT results for all these NNP models, 
and the NNP models can also predict the formation energy once SrTiO$_3$:V$_{\text{O}}$ data are included in the training data sets. 
Even for larger supercell sizes that are difficult to calculate using first-principles calculations, the formation energies evaluated from the NNP-MD simulations well reproduce the extrapolated DFT values. 
This study offer important knowledge for constructing accurate NNP models to describe point-defect systems including SrTiO$_3$:V$_{\text{O}}$. 

\end{abstract}


\section{Introduction}

Oxygen vacancies (V$_{\text{O}}$) in perovskite oxide SrTiO$_3$ have been studied as a typical example of anion defects in perovskite compounds. 
Although pristine SrTiO$_3$ is considered an intrinsic semiconductor,~\cite{Piskunov_2004,Begum_2019} 
the synthesized samples often exhibit n-type semiconducting behaviors due to the presence of V$_{\text{O}}$. 
These oxygen defects make a significant effect on the electronic structure of SrTiO$_3$ due to the formation of defect levels within the energy band gap 
since V$_{\text{O}}$ acts as an n-type doping and introduces excess electron carriers.~\cite{Hou_2010,Samanta_2012}  
The static electronic properties, including the density of states, band structure, and the formation energy of V$_{\text{O}}$, 
have been investigated using first-principles calculations based on density functional theory (DFT). 
However, the dynamics and thermodynamics, such as oxygen vacancy diffusion and heat capacity, are difficult to address using molecular dynamics (MD) simulations: 
ab initio MD (AIMD) simulations are restricted to small-scale systems and short-time simulations due to their high computational costs, 
while the accuracy of classical MD simulations is limited despite their low computational costs.~\cite{Katsumata_1998,Schie_2012,Waldow_2016,Dawson_2020} 

In recent decades, machine learning (ML) interatomic potentials, including neural network potential (NNP),~\cite{NNP_2007} 
Gaussian approximation potential (GAP),~\cite{GAP_2010} and spectral neighbor analysis potential (SNAP),~\cite{SNAP_2015}  
have attracted much interest because of their lower computational costs by several order of magnitude while achieving comparable accuracy with DFT calculations. 
These ML potentials have been successfully applied not only to molecules and solids but also to complex systems 
such as alloy,~\cite{Shimizu_2021} amorphous materials,~\cite{Andrade_2020} and interfaces.~\cite{Chaney_2024} 
MD simulations with the NNP model (NNP-MD) have also been applied to pristine SrTiO$_3$ to describe the structural phase transition between the cubic and tetragonal crystal phase.~\cite{He_2022} 
Currently, active learning (AL) techniques are being actively developed to reduce the amount of labeled data.~\cite{Smith_2018,Vandermause_2020,Sivaraman_2020,Zhang_2020,Shimizu_2021,Novikov_2021,Podryabinkin_2023,van_der_Oord_2023,Kulichenko_2023} 
The performance of NNP models typically depends on the quality and size of the training data sets; 
however, the data labeling process to calculate the energy and atomic forces is computationally very expensive. 
AL processes iteratively add diverse structures to the training data, focusing on regions of the configuration space where the NNP model exhibits poor predictive accuracy. 

This paper presents NNP-MD simulations of 
SrTiO$_3$ with and without oxygen vacancies to predict their physical quantities, including the total energy of pristine SrTiO$_3$ and formation energy of SrTiO$_3$:V$_{\text{O}}$. 
We make use of an NNP generator, GeNNIP4MD (Generator of Neural Network Potential for Molecular Dynamics),~\cite{Matsumura_2025} to generate NNP models available for NNP-MD simulations. 
Several NNP models are generated on different data sets to verify their accuracy,  
and all these NNP models accurately predict the total energy evaluated from DFT. 
The formation energy can also be reproduced when SrTiO$_3$:V$_{\text{O}}$ data are included in the training data sets. 
In particular, even for larger supercell sizes that are difficult to calculate using first-principles calculations due to the high computational cost, 
the formation energies evaluated from the NNP-MD simulations well reproduce the extrapolated DFT values. 
These verifications suggest that the NNP models that accurately describe pristine SrTiO$_3$ and SrTiO$_3$:V$_{\text{O}}$ are constructed by the training procedures and the NNP-MD simulations can be performed with sufficient accuracy. 

\section{Methodology}

The total energy of pristine SrTiO$_3$ and formation energy of SrTiO$_3$:V$_{\text{O}}$ predicted by the NNP-MD simulations are compared with the DFT results. 
The NNP models are constructed by the training data set of the AIMD simulations using the $2 \times 2 \times 2$ and/or $3 \times 3 \times 3$ supercell models. 
The total energy and formation energy of the $2 \times 2 \times 2$, $3 \times 3 \times 3$, $4 \times 4 \times 4$, and $5 \times 5 \times 5$ supercell models evaluated by the NNP-MD simulations are verified using the DFT results of the $2 \times 2 \times 2$, $3 \times 3 \times 3$, and $4 \times 4 \times 4$ supercell models and the extrapolated value of the $5 \times 5 \times 5$ one. 

\subsection{AIMD based on DFT}

\begin{figure*}
\includegraphics[width=15.0cm,clip]{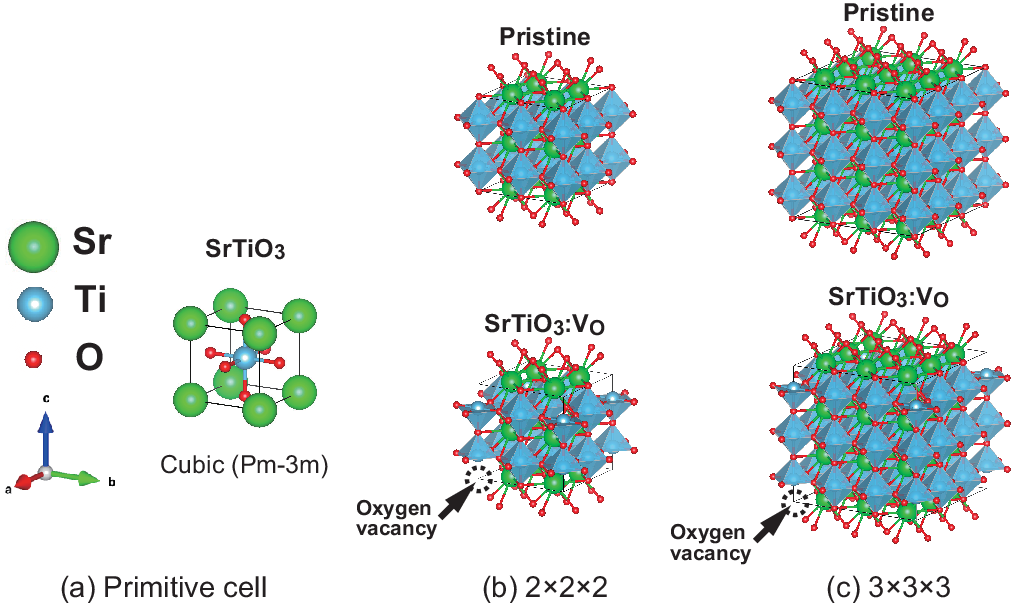}
\caption{
(a) Crystal structure of SrTiO$_3$. (b) $2 \times 2 \times 2$ and (c) $3 \times 3 \times 3$ supercell models with and without a single oxygen vacancy (i.e., SrTiO$_3$:V$_{\text{O}}$ and pristine SrTiO$_3$). 
The primitive cell of SrTiO$_3$ contains 5 atoms. 
The $2 \times 2 \times 2$ supercell models contain 40 atoms for pristine SrTiO$_3$ and 39 atoms for SrTiO$_3$:V$_{\text{O}}$. 
The $3 \times 3 \times 3$ supercell models contain 135 atoms for pristine SrTiO$_3$ and 134 atoms for SrTiO$_3$:V$_{\text{O}}$. 
The TiO$_6$ octahedral is highlighted in blue for (b) and (c).
}
\label{fig:supercell}
\end{figure*}

To generate an initial data set for NNP construction, DFT-based AIMD calculations are performed using supercells of pristine SrTiO$_3$ and SrTiO$_3$ in the presence of oxygen vacancies. 
Figure~\ref{fig:supercell} schematically shows these supercells of different sizes: $2 \times 2 \times 2$ and $3 \times 3 \times 3$ supercells of SrTiO$_3$ with and without a single oxygen vacancy, 
namely, SrTiO$_3$:V$_{\text{O}}$ and pristine SrTiO$_3$. 
The $2 \times 2 \times 2$ ($3 \times 3 \times 3$) supercell models of pristine SrTiO$_3$ and SrTiO$_3$:V$_{\text{O}}$ contain 40 and 39 (135 and 134) atoms, respectively. 
The AIMD simulations are performed applying a constant NVT ensemble with 2 fs time steps for 1 ps (500 steps). 
Here, a Nose-Hoover thermostat with a Nose-mass parameter of 0.1 is used 
and the initial temperature is set to 300 K and linearly increased up to 1300 K (at 1.0 K/fs) to obtain the various random structures. 
Further structural generations are carried out using NNP-based MD simulations due to the high computational costs of AIMD calculations, 
and subsequent static DFT calculations are performed for the extracted structures by the sampling and screening method.

These AIMD calculations are performed using the Vienna Ab initio Simulation Package (VASP)~\cite{Kresse_1993,Kresse_1994,Kresse_1996a,Kresse_1996b} 
with the projector augmented wave (PAW) method~\cite{Blochl_1994,Kresse_1999} within the Perdew--Burke--Ernzerhof (PBE) exchange-correlation functional~\cite{PBE_1996}.
The simplified Hubbard {\it U} correction (PBE+{\it U}) of Dudarev {\it et al.}~\cite{Dudarev_1998} is applied to the Ti-3$d$ orbitals with $U -J = 4.36$ eV.~\cite{Kim_2009,Cuong_2007,Choi_2013,Hou_2010} 
For PAW potentials, the core orbitals (number of valence electrons) are set to be Sr:[Ar]3$d^{10}$ (10), Ti:[Ne] (12), O:[He] (6). 
The plane-wave cutoff energy of 500 eV and the Gaussian smearing with the smearing width of 0.01 eV are used, 
and 
the NVT simulations are performed fixing the lattice constants at the experimental values ($|\bm{a}_{1,2,3}| = 3.905$ \AA \, for STO.~\cite{Madsen_1995,Ligny_1996,Cao_2000,Shimuta_2002})
The Monkhorst-Pack $12 \times 12 \times 12$, $6 \times 6 \times 6$, $4 \times 4 \times 4$, and $3 \times 3 \times 3$ $k$-point grids are used for the primitive cell, the $2 \times 2 \times 2$, $3 \times 3 \times 3$, and $4 \times 4 \times 4$ supercell models, respectively. 

\subsection{Neural network potential generation}

\begin{figure*}
\includegraphics[width=8.0cm,clip]{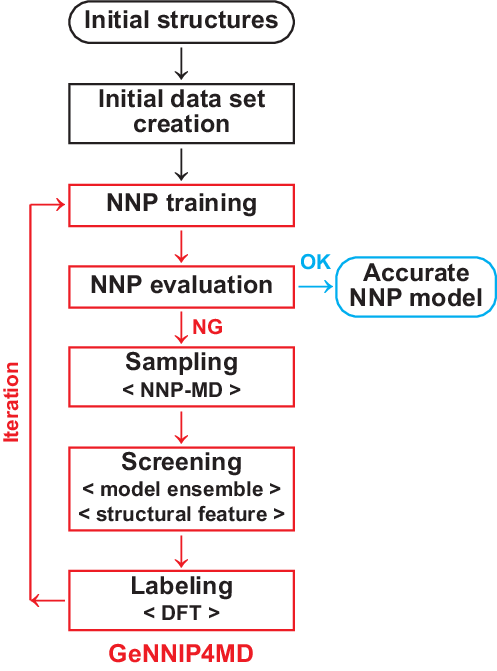}
\caption{
A flowchart of the NNP generation in GeNNIP4MD. 
}
\label{fig:gennip4md}
\end{figure*}

A NNP generator GeNNIP4MD is used to generate NNP models available for NNP-MD simulations. 
GeNNIP4MD also allows for efficient sampling and screening methods for the initial structure inputs through AL process. 
The data set is composed of structures along with their corresponding energies, forces, and stresses calculated using DFT. 
After an initial data set creation phase, the AL workflow of GeNNIP4MD proceeds through the following sequential phases: NNP training, NNP evaluation, sampling, screening, and labeling (Figure~\ref{fig:gennip4md}). 

\subsubsection{Initial data set creation}

\begin{figure*}
\includegraphics[width=15.0cm,clip]{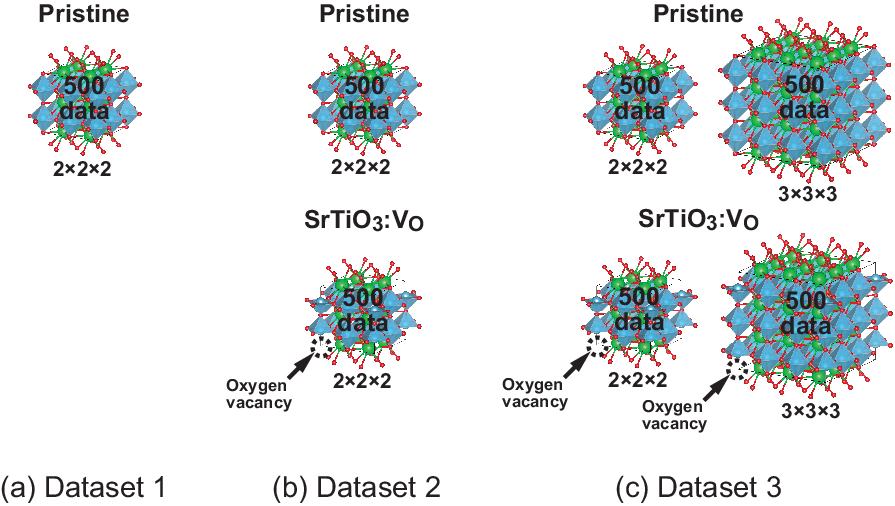}
\caption{
Contents of trained data set. 
(a) Dataset1 includes 500 data points, (b) Dataset2 1,000 data points, and (c) Dataset3 2,000 data points. 
500 AIMD trajectories are prepared for each different size of the supercell models with/without a single oxygen vacancy. 
}
\label{fig:dataset}
\end{figure*}

AL processes begin with the creation of an initial data set to train an initial NNP model. 
The generator can create the initial data set using AIMD simulations with the Atomic Simulation Environment (ASE) package.~\cite{Larsen_2017} 
The initial data sets of pristine SrTiO$_3$ and SrTiO$_3$:V$_{\text{O}}$ are created by GeNNIP4MD using the AIMD simulations under the calculation conditions mentioned in the previous section. 
We here prepare different initial data sets from the different supercell models to compare the accuracy of the NNP models: 
Dataset1, 2, and 3 contain the AIMD trajectories of the $2 \times 2 \times 2$ supercell model without V$_{\text{O}}$ (500 data points), 
those of the $2 \times 2 \times 2$ supercell model with and without V$_{\text{O}}$ (1,000 data points), 
and the $2 \times 2 \times 2$ and $3 \times 3 \times 3$ supercell models with and without V$_{\text{O}}$ (2,000 data points), respectively (Figure~\ref{fig:dataset}). 

\subsubsection{NNP training}

The NNP model is trained on the data set in this phase. 
Several deep potential (DP) models with an identical architecture but with different initialization seeds are trained simultaneously to estimate the uncertainty criteria that is a description of the screening phase. 
The DP model initialized with the first seed is employed in the subsequent evaluation and sampling phases, while all models are used in the screening phase. 
Here, the NNP models are constructed using the DeepPot-SE~\cite{Zhang_2018} implemented in DeePMD-kit,~\cite{Wang_2018,Zeng_2023}
and four DP models are used with 500,000 training steps. 

\subsubsection{NNP evaluation}

This phase evaluates the trained NNP model to verify its accuracy and reliability. 
The energy and force obtained from the NNP model are compared to DFT results using the mean absolute error (MAE) and the root mean squared error (RMSE). 

\subsubsection{Sampling}

In this phase, NNP-MD simulations using the trained NNP model are performed to explore the configuration space and generate a set of candidate structures.
The Large-scale Atomic/Molecular Massively Parallel Simulator (LAMMPS) package~\cite{Thompson_2022} is used for the NNP-MD simulations. 
By leveraging these NNP-MD simulations, the NNP generator can effectively sample a wide range of configurations, ensuring that the final NNP model is well-trained and capable of accurate predictions under the specified target conditions.
Here NNP-MD simulations in the constant temperature, constant volume (NVT) are performed, allowing for the exploration of the configuration space by varying the temperatures (300--2400 K). 

\subsubsection{Screening}

This phase reduces the number of candidate structures passed to the labeling phase. 
To improve the current NNP model, the candidate structures that are not present in the existing data set are identified using a combination of two screening methods: 
model ensemble-based~\cite{Smith_2018,Zhang_2020} and structural feature-based~\cite{Shimizu_2021,Hajinazar_2017,Cubuk_2017} methods. 
In the NNP generator, the model ensemble-based screening method is initially applied to filter candidates based on model prediction accuracy, 
and the structural feature-based screening method is further used to refine the selection based on structural features. 
The former method reduces the number of candidates, while the latter narrows it down further to a manageable number.
We here make the former method left candidate structures with forces in the range of 0.05 to 0.20 eV/\AA, and the latter method left 1/5 of the number of the initial data set.

\subsubsection{Labeling}

In this phase, the reference energies, forces, and stresses for the selected candidate structures are evaluated from DFT calculations. 
The labeled structures are further filtered based on a threshold for the absolute magnitude of the maximum force, 
and  the remaining labeled structures are split into training and validation data sets according to a ratio (1:4). 
The AL workflow of GeNNIP4MD returns to the NNP training phase until a NNP model with the desired accuracy can be obtained. 
Here the AL processes are limited to 5 iterations, through which additional 500 and 1,000 data (namely, 100 and 200 data per iteration) are taken into account for Dataset1 and 2, respectively. 

\subsection{NNP-MD simulation}

NNP-MD simulations are carried out using the NNP models obtained from the NNP construction described above. 
Each NNP model is set to be the DP model with the first seed in four DP ones. 
When the NNP-MD predictions are compared to the DFT results, the NNP-MD simulations are performed in a constant NVT ensemble with 1 fs time steps for 10 ps (10,000 steps) at zero temperature. 
This MD simulation corresponds to the structural optimization of atomic positions fixing lattice constants. 

\section{Results and Discussion}

The NNP-MD predictions of the total energy of pristine SrTiO$_3$ and the formation energy of SrTiO$_3$:V$_{\text{O}}$ are compared to the DFT results after the accuracy is verified for different NNP models. 
These different NNP models constructed from Dataset1, 2, and 3 are denoted as NNP1, 2, and 3. 
The DFT results are obtained from the structural optimization of the atomic positions until the residual Hellmann-Feynman force becomes less than 0.01 eV/\AA. 

\subsection{Accuracy}

\begin{figure*}
\includegraphics[width=17.0cm,clip]{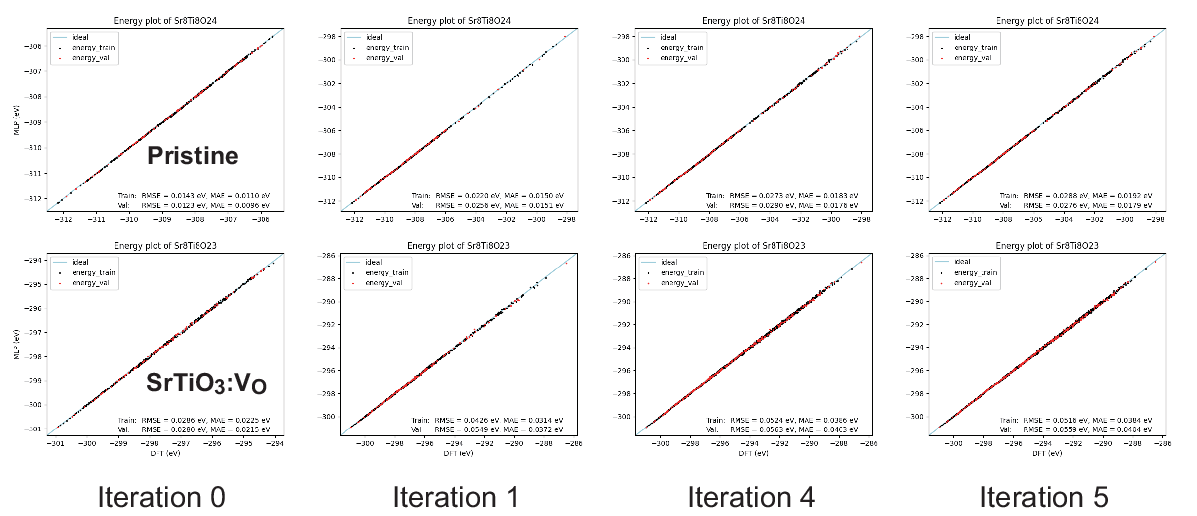}
\caption{
Parity plot of MLP and DFT energies in NNP2 at the iteration 0, 1, 4, and 5. 
}
\label{fig:energy}
\end{figure*}

\begin{figure*}
\includegraphics[width=17.0cm,clip]{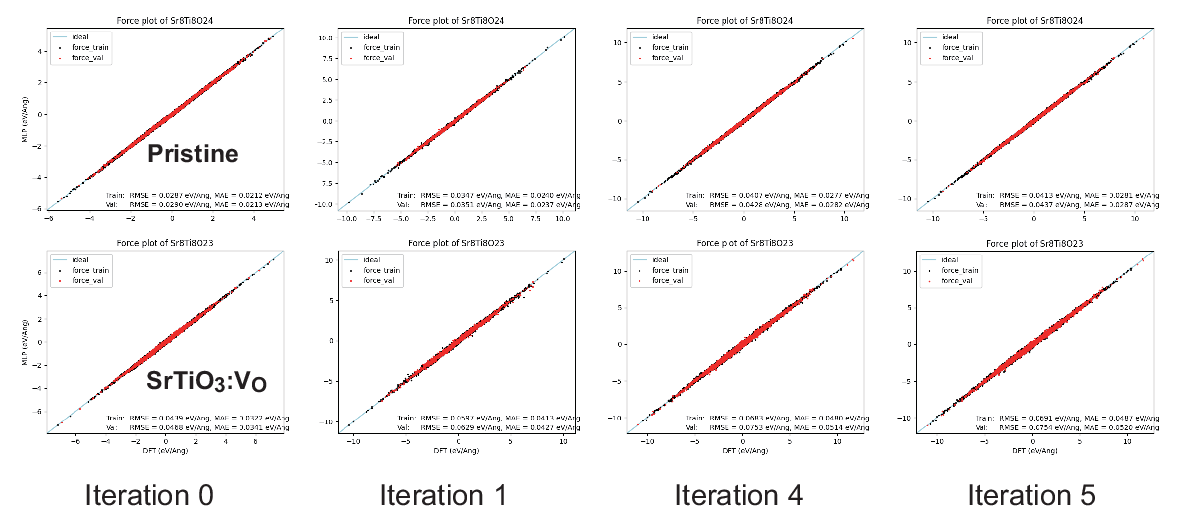}
\caption{
Parity plot of MLP and DFT forces in NNP2 at the iteration 0, 1, 4, and 5. 
}
\label{fig:force}
\end{figure*}

To see the accuracy of the NNP models, the ML potential (MLP) predictions and DFT results are compared using the RMSE and MAE. 
Figure~\ref{fig:energy} (\ref{fig:force}) shows the parity plot of the MLP and DFT energies (forces) for pristine SrTiO$_3$ and SrTiO$_3$:V$_{\text{O}}$ in NNP2, respectively. 
The accuracy of the iteration 0, 1, 4, and 5 is displayed to examine the iterative dependence throughout the AL processes. 
Since the RMSE and MAE of the energies (forces) are less than 60 meV (80 meV/\AA) for each iteration, the NNP models are considered to be well-trained for each AL process. 
We can also find that the RMSE and MAE tend to increase immediately after the start of the AL processes, and then saturate or slightly decrease at the end of the iteration. 
This is because the accuracy is temporarily reduced by adding structures with larger energy and force through the AL processes. 

\subsection{Total energy}

\begin{figure*}
\includegraphics[width=7.5cm,clip]{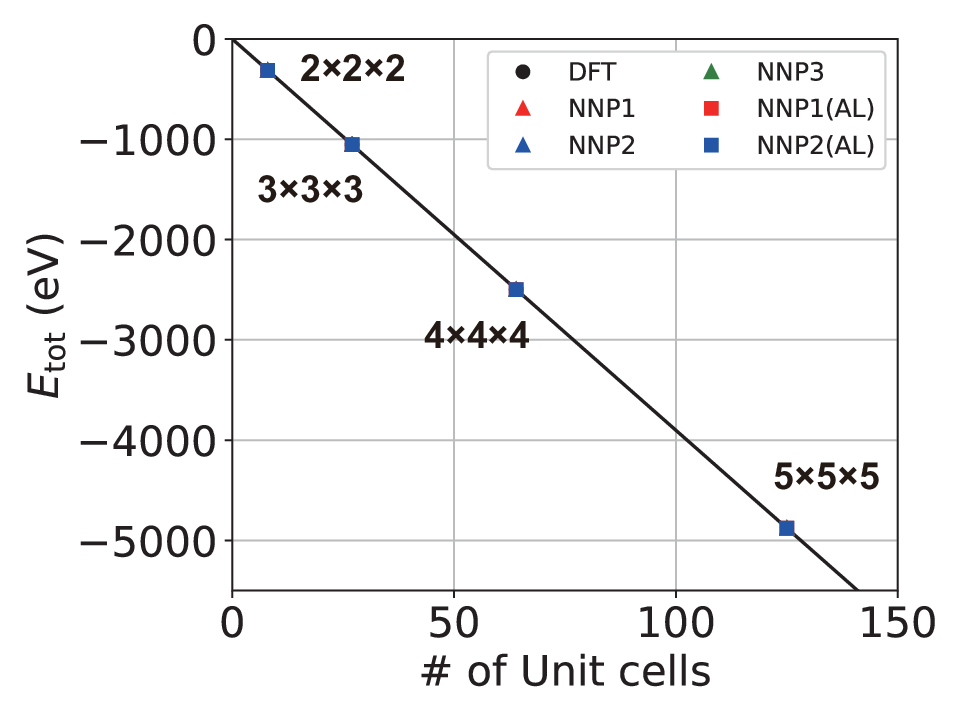}
\caption{
NNP-MD prediction of the total energy of the pristine SrTiO$_3$ for $2 \times 2 \times 2$, $3 \times 3 \times 3$, $4 \times 4 \times 4$ and $5 \times 5 \times 5$ supercell sizes. 
The black line is the linear fit of the DFT total energy for the $2 \times 2 \times 2$, $3 \times 3 \times 3$, and $4 \times 4 \times 4$ supercell sizes. 
}
\label{fig:Etot}
\end{figure*}

Figure~\ref{fig:Etot} shows the NNP-MD prediction of the total energy of the pristine SrTiO$_3$ for different supercell sizes. 
The predicted total energy of $2 \times 2 \times 2$, $3 \times 3 \times 3$, $4 \times 4 \times 4$, and $5 \times 5 \times 5$ supercell models are compared to the DFT results for $2 \times 2 \times 2$, $3 \times 3 \times 3$, and $4 \times 4 \times 4$ ones. 
Regardless of the training data, each NNP model accurately predicts the DFT total energies of integer multiples of the primitive cell of the pristine SrTiO$_3$.
The total energy of larger supercell sizes ($4 \times 4 \times 4$ and $5 \times 5 \times 5$) than those included in the training data can also be evaluated using these NNP models. 

\subsection{Formation energy}

The formation energy of SrTiO$_3$:V$_{\text{O}}$ can be expressed as~\cite{Zhang_1991,Komsa_2012,Freysoldt_2014,Oba_2018} 
\begin{equation}
E_{\text{f}} = E[\text{SrTiO$_3$:V$_\text{O}$}] -E[{\text{SrTiO$_3$}}] -E_{\text{O}} , 
\end{equation}
where $E[\text{SrTiO$_3$:V$_\text{O}$}]$ is the total energy of the supercell with a single V$_{\text{O}}$, $E[{\text{SrTiO$_3$}}]$ is the total energy of the pristine supercell, 
and $E_{\text{O}}$ denotes the total energy of an oxygen atom. 
Here $E_{\text{O}}$ is set to be half the DFT total energy of O$_2$ molecule in order to focus on the accuracy of the formation energy. 

\begin{figure*}
\includegraphics[width=7.5cm,clip]{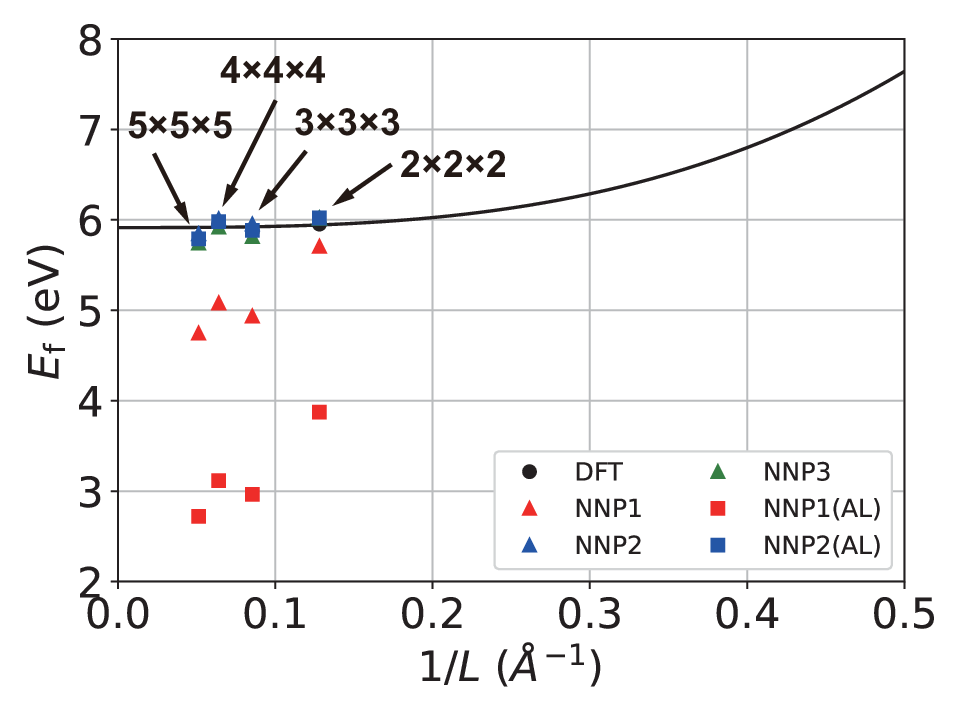}
\caption{
NNP-MD prediction of the formation energy of SrTiO$_3$:V$_{\text{O}}$ for $2 \times 2 \times 2$, $3 \times 3 \times 3$, $4 \times 4 \times 4$, and $5 \times 5 \times 5$ supercell sizes. 
The black line is the extrapolation of the DFT formation energy for the $2 \times 2 \times 2$, $3 \times 3 \times 3$, and $4 \times 4 \times 4$ supercell sizes. 
}
\label{fig:Ef}
\end{figure*}

The NNP-MD prediction of the formation energy of SrTiO$_3$:V$_{\text{O}}$ for different sizes is shown in Figure~\ref{fig:Ef}. 
The predicted formation energy of $2 \times 2 \times 2$, $3 \times 3 \times 3$, $4 \times 4 \times 4$, and $5 \times 5 \times 5$ supercell models are compared to the DFT results for $2 \times 2 \times 2$, $3 \times 3 \times 3$, and $4 \times 4 \times 4$ ones. 
The DFT extrapolation of the formation energy in the dilute limit is also depicted: 
the finite-size correction for a cubic (isotropic) supercell can be given as a function of the linear dimension of the cubic supercell $L$,~\cite{Leslie_1985,Makov_1995,Fraser_1996,Dabo_2008} and the leading contribution depends on $1/L^{3}$ for neutral states. 
NNP2 and 3 can reproduce the formation energy obtained from DFT for all different supercell sizes. 
These NNP models are able to predict the formation energies of larger supercell models ($4 \times 4 \times 4$ and $5 \times 5 \times 5$) not included in the data set. 
In particular, the extrapolated value for the $5 \times 5 \times 5$ supercell model, which is difficult to calculate using DFT, can be evaluated with high accuracy. 
We can also find that only NNP1 fails to predict the formation energy for each size, in contrast to the predictions of the total energy. 
This is because the data set used to train NNP1 (Dataset1) does not contain any configurations of SrTiO$_3$:V$_{\text{O}}$. 

\section{Conclusion}

The NNP-MD simulations of pristine SrTiO$_3$ and SrTiO$_3$:V$_{\text{O}}$ were performed to predict the physical quantities, including the total energy and the formation energy of SrTiO$_3$:V$_{\text{O}}$. 
The NNP models were generated on different data sets to verify their accuracy by comparing the NNP-MD predictions with DFT results. 
For all these NNP models, the predictions of the total energy for different supercell sizes were in good agreement with the DFT total energies.
The formation energy can also be reproduced for different supercell sizes when data from SrTiO$_3$:V$_{\text{O}}$ were included in the data sets. 
In particular, such NNP models were able to accurately predict the extrapolated DFT value for the larger supercell model that is difficult to calculate for first-principles calculations. 
This finding will contribute to the understanding and development of large-scale and long-time NNP-MD simulations of functional materials with point defects.

%
%

\begin{acknowledgement}

K.N. wishes to thank Yuto Iwasaki (Fujitsu) for fruitful discussions. 
A part of the numerical calculations were performed using the supercomputing system at the Institute for Solid State Physics (ISSP) Supercomputer Center of the University of Tokyo.  

\end{acknowledgement}

%
%
%


\bibliography{refs}

\end{document}